# A multi-channel real-time bioimpedance measurement device for pulse wave analysis

Roman Kusche, Paula Klimach, and Martin Ryschka

*Abstract—* Pulse wave analysis is an important method used to gather information about the cardiovascular system. Instead of detecting the pulse wave via pressure sensors, bioimpedance measurements can be performed to acquire minuscule changes in conductivity of the tissue, caused by the pulse wave. This work presents a microcontroller based bioimpedance measurement system, which has the capability to acquire impedance measurements from up to four independent channels simultaneously. By combining a problem-specific analog measurement circuit with a 24 bits ADC, the system is capable of acquiring 1000 impedances per second with a signal to noise ratio in a range from 92 dB to 96 dB. For data storage and analysis the digitized data are sent via USB to a host PC. A graphical user interface filters and plots the data of all channels in real-time. The performance of the system regarding measuring constant impedances, as well as impedance changes over time is demonstrated. Two different applications for pulse wave analysis via multi-channel bioimpedance measurements are presented. Additionally, first measurement results from a human subject are shown to demonstrate the system's applicability of analyzing the pulse wave morphology as well as the aortic pulse wave velocity.

## I. INTRODUCTION

Pulse wave analysis is a biomedical technique to get information about arterial stiffness [1]-[3]. In the past, especially tonometric or oscillometric methods were used to detect the pressure pulse waves, in the arteries, which are generated by the pumping of the heart [4]-[7]. The most significant parameters are the pulse wave velocity (PWV) inside the aorta and the morphology change of the pulse wave inside the aorta [8]-[12]. Since most non-invasive techniques are not able to detect the pulse wave directly at the aortic arch or at the end of the aorta, other measurement setups were developed to circumvent this problem. These measurement setups detect the pulse wave at other positions of the human body, like at the extremities, mostly using cuff-based pressure measurements. Afterwards, transfer functions are used to estimate the pressure conditions inside the aorta [5], [13].

Instead of performing pressure measurements, the pulse wave can also be acquired by bioimpedance measurements, called impedance plethysmography [14]-[17]. This technique is based on the fact, that blood has a higher conductivity than other tissue, such as fat or muscle, in the frequency range between 10 kHz and 1 MHz [17]. Since the volume of blood in the arteries increases, when the arriving pressure pulse wave widens the arteries, the bioimpedance magnitude decreases within an area under test. The advantage of this non-invasive technique is that arteries, even deep under the skin surface, can be analyzed. Additionally, the subject's discomfort, caused by sensors pressed onto the skin above the arteries, is replaced by more comfortable electrodes.

A challenge of the impedance plethysmography is the very low impedance change in m$\Omega$ ranges [14], [16]. Due to reasons of electrical safety, the excitation currents of the bioimpedances are strictly limited. Useful excitation current frequencies are between 10 kHz and 500 kHz. Although the conductivity of blood is frequency dependent [17]-[18], measurements have proven that the choice of the excitation frequency does not significantly change the morphology of the acquired pulse wave [14]. Since the numerical value of the bioimpedance and its variations is not of major interest in this application, its frequency dependency can be neglected as well.

Bioimpedance measurement devices, described in other publications are usually not focused on synchronized multi-channel measurements and high accuracy measurements at the same time. Descriptions of single channel systems with sufficient accuracies to acquire pulse waves in the tissue, already exist [14], [16], [19]. In reference to multi-channel systems [20]-[24], the major application is the electrical impedance tomography (EIT). But none of these systems is intended to acquire the pulse wave at different positions of the human body simultaneously and to display the signals in real-time.

The introduced system is capable of measuring four bioimpedances simultaneously. Since for the pulse wave

Manuscript received xxx xx, 201x; revised xxx xx, 201x and xxx xx, 201x; accepted xxx xx, 201x. Date of publication xxx xx, 201x; date of current version xxx xx, 201x. This publication is a result of the ongoing research within the LUMEN research group, which is funded by the German Federal Ministry of Education and Research (BMBF, FKZ 13EZ1140A/B).
R. Kusche and M. Ryschka are with the Laboratory of Medical Electronics (LME), Luebeck University of Applied Sciences, 23562 Luebeck, Germany. (email: roman.kusche@fh-luebeck.de; martin.ryschka@fh-luebeck.de).
P. Klimach is with the Laboratory of Medical Electronics (LME), Luebeck University of Applied Sciences and with the Graduate School for Computing in Medicine and Life Sciences, University of Luebeck, 23562 Luebeck, Germany. (email: paula.klimach@fh-luebeck.de)

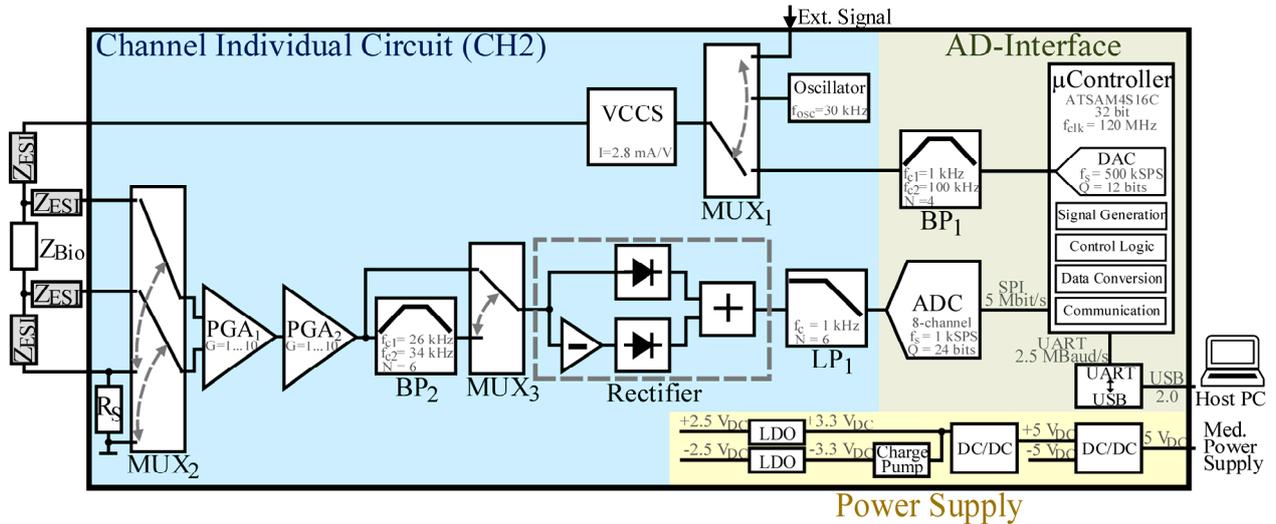

Fig. 1. Principle block diagram of the microcontroller based bioimpedance measurement device. For better visualization, the signal path of just one channel (CH2) is shown in detail. The Channels 2…4 are identically realized. The block diagram is separated into the analog-digital (AD) interface (green), the power supply section (yellow) and the channel individual measurement circuit (blue). The DAC (digital-to-analog converter) of the microcontroller generates a sinusoidal voltage signal, which is filtered by the band pass $BP_1$. Multiplexer $MUX_1$ can be used to select one of three excitation signal sources. The voltage-controlled current source (VCCS) converts the voltage signal into a current, which is applied via the electrode skin impedances ($Z_{ESI}$) to the bioimpedance ($Z_{Bio}$) and passes a shunt resistor $R_S$. The occurring voltage drop can be amplified by two programmable gain amplifiers (PGA) and is processed by an optional band pass filter ($BP_2$), an analog rectifier circuit and a low pass filter ($LP_1$). After digitizing the signal, the data are sent via the serial peripheral interface (SPI) to the microcontroller. Finally, a Universal Asynchronous Receiver Transmitter (UART) to Universal Serial Bus (USB) interface chip transmits the data to a host PC.

analysis, only the magnitude of the bioimpedance is necessary, the phase shifts are not measured by the system. The combination of an analog voltage measurement circuit and a 24 bits analog-to-digital converter (ADC) enables impedance measurements with signal-to-noise ratios of up to 96 dB. Because of digitizing the bioimpedance values directly with a sample rate of 1000 samples per second (SPS), the data transfer rate to a host PC is much lower than that of raw data acquiring systems and the digital signal processing can be very elementary. Real-time filtering and plotting is performed by a developed graphical user interface (GUI) on a host PC.

This work describes the development and the verification of the measurement system, as well as two exemplary measurements for pulse wave analysis.

## II. MATERIALS AND METHODS

### A. System architecture

In Fig. 1, a principle block diagram of the microcontroller based bioimpedance measurement system is shown. The system determines the bioimpedance under test ($Z_{Bio}$) on the left side and communicates with a host PC on the right side. For better display, the implementation of just one impedance measurement channel (CH2, blue) is illustrated, whereat the remaining three channels are implemented identically. The green analog-digital (AD) block, as well as the yellow power supply block are implemented just once and are shared by the measurement blocks.

The sinusoidal excitation signal can be generated by the microcontroller's (ATSAM4S16C, Atmel, San José, CA, US) internal digital-to-analog converter (DAC) with a sample rate of $f_s$=500 kSPS and a resolution of Q=12 bits. After passing a bandpass filter ($BP_1$) for reconstruction and DC removal, the signal is connected to a 3-to-1 multiplexer ($MUX_1$) input. This multiplexer can be used to select one of three excitation signals including the signal from the aforementioned DAC, an analog oscillator or an external voltage. The selected voltage signal is converted by an AD8130 (Analog Devices, Analog Devices, Norwood, MA, US) based enhanced voltage controlled current source (VCCS) circuit [25] into an alternating current (AC). In order to take measurements of the current, a shunt resistor ($R_S$) is connected in series to the electrode skin impedances ($Z_{ESI}$) and the bioimpedance of interest ($Z_{Bio}$).

Either the voltage over the bioimpedance or the voltage across the shunt resistor can be chosen for measurement by using the multiplexer $MUX_2$ (ADG1236, Analog Devices, Norwood, MA, US). The selected signal can be amplified by two programmable gain amplifiers (PGA, AD8250, Analog Devices, Norwood, MA, US), which are connected successively, in 9 fixed steps (G={1; 2; 4; 5; 10; 20; 25; 50; 100}). A multiplexer ($MUX_3$, TS12A12511, Texas Instruments, Dallas, TX, US) can be used to choose between a pre-filtered signal and the direct PGA output for the next signal processing steps. The filter is set to have different frequency responses in each channel and is especially useful in multi-channel applications to differentiate between the channels. In the exemplary illustrated block diagram of channel 2, this individual filter is implemented as a 6th order band pass ($BP_2$) whose frequency response is intended to let the signal of the signal generation oscillator ($f_c$=30 kHz) pass.

Afterwards, the sinusoidal signal is rectified by an analog switched full-wave rectifier circuit, which is designed to process low voltage signals with high frequencies in a range from 10 kHz to 500 kHz. This rectifier consists mainly of two identical half-wave rectifier circuits and is described in section C in detail.

To extract the information about the impedance's magnitude, the rectified signal is filtered by an active low pass

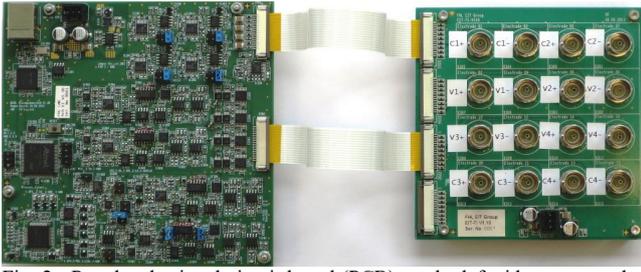

Fig. 2. Populated printed circuit board (PCB) on the left side, connected to an interface board. The measurement PCB's dimensions are about 128 mm x 119 mm and it contains 813 components.

($LP_1$) of $6^{th}$ order with a cut-off frequency of 1 kHz, realized with the LMV844 (Texas Instruments, Dallas, TX, US) in Multiple Feedback topology with Butterworth characteristic.

For digitizing the analog output signals of all four impedance channels, an 8-channel analog-to-digital converter (ADS131E08, Texas Instruments, Dallas, TX, US) is used. This ADC digitizes all channels synchronously with a resolution of Q=24 bits and a sample rate of $f_s$=1 kSPS. The remaining four ADC channels are used for acquisition of other bio signals like Electrocardiogram or Photoplethysmogram, which are not further described in this work.

The digitized data are transmitted to the microcontroller via the serial peripheral interface (SPI). After data conversions, the microcontroller transmits the measurement results via a Universal Asynchronous Receiver Transmitter (UART) to a Universal Serial Bus (USB) interface chip (FT2232HL, Future Technology Devices International, Glasgow, UK). To comply with the IEC-60601-1, the USB connection between this chip and the host PC is isolated by an optical interface device (USB-GT-MED-D, Meilhaus Electronic, Alling, DE).

A graphical user interface can be used to analyze the measurement data and to change the setup of the hardware device in real-time. Furthermore, a MATLAB (The MathWorks, Natick, MA, US) interface can also be used for specific measurement protocols.

The bioimpedance measurement system is powered with $\pm 5$ $V_{DC}$ by an external medical power supply (MPU31-102, SINPRO Electronics, Pingtung City, TW) in combination with a DC/DC converter (LT8582, Linear Technology, Milpitas, CA, US). Additionally, $\pm 3.3$ $V_{DC}$ and $\pm 2.5$ $V_{DC}$ for the digital components and the ADC are generated internally by an additional DC/DC converter (TPS63001, Texas Instruments, Dallas, TX, US), a charge pump (SP6661, MaxLinear High Performance Analog, San Jose, CA, US) and two low drop-out regulators (TPS73025, TPS72325, Texas Instruments, Dallas, TX, US), as shown in the block diagram. The power consumption of the system, using all four impedance channels is between 4.9 W and 5.7 W, depending on the PGA setups and the load impedance. A photograph of the printed circuit board (PCB), connected to an additional measurement cable interface board is shown in Fig. 2. To avoid coupling effects on the PCB, differential pair routing has been applied to the signal paths.

### B. Principle of bioimpedance measurement

In contrast to other recently published bioimpedance measurement systems [27], [14], the described system's measurement principle is based on analog signal processing. The major advantage of analog pre-processing, more specifically of analog demodulation, is that digitizing the high frequency bioimpedance carrier frequency is not necessary. Hence, the ADC sampling rate has to be just high enough to digitize the useful signal, which is the actual impedance magnitude information. For these low sampling frequency ranges, very high resolution ADCs are available in the market.

During a typical bioimpedance measurement with the described system, a 2 mA AC current with a frequency between 10 kHz and 500 kHz is injected via the current electrodes into the tissue of interest. By using an external signal source, smaller current amplitudes can be chosen.

At the beginning of each measurement, the actual current is determined by measuring the voltage drop over the known shunt resistor. Afterwards, the multiplexer connects the bioimpedance with the voltage measurement input. Since it can be assumed, that the output impedance of the current source is much higher than the load impedance and its changes during a measurement procedure, a single current measurement is sufficient and just the voltage drop across the bioimpedance is measured over time. In special cases, where the precise current shall also be measured, a second channel can be used for this purpose.

The combination of two PGAs within the voltage measurement signal path enables for a wide range of different gains {G=1; 2; 4; 5; 10; 20; 25; 50; 100} to adjust the corresponding impedance measurement ranges {$Z_{max.}$=1180 Ω; 590 Ω; 295 Ω; 236 Ω; 118 Ω; 59 Ω; 47.2 Ω; 23.6 Ω; 11.8 Ω}.

The ADC's sample rate is adjustable and chosen to be 1 kSPS, thus corresponding to a measurement rate of 1000 impedances per second and per channel. According to the ADC's datasheet this condition leads to the effective number of bits (ENOB) of 19.6 bits [26].

### C. Switched high-frequency rectifier

The switched rectifier circuit is designed to work for frequencies of up to 500 kHz. Illustrated in Fig. 3 is the block diagram of the designed electronic circuit. At first, the original sinusoidal signal $V_{in}$ is split and one part is inverted by the inverting amplifier $IC_2$ (OPA2727, Texas Instruments, Dallas, TX, US). The buffer $IC_1$ is implemented to avoid additional undesired phase shifts between both signal paths, caused by the components characteristics. $IC_3$ (OPA2690, Texas Instruments, Dallas, TX, US) and $IC_4$ are configured as complementary comparators to detect the positive and the negative signal half-wave, respectively. Their resulting rectangular outputs control the gates of the N-channel transistors $T_1$ (PMDT290UNE, NXP Semiconductors, Eindhoven, NL) and $T_2$. To combine both the half-wave rectified output signals, a summing amplifier $IC_5$ (LMH6628, Texas Instruments, Dallas, TX, US) is implemented.

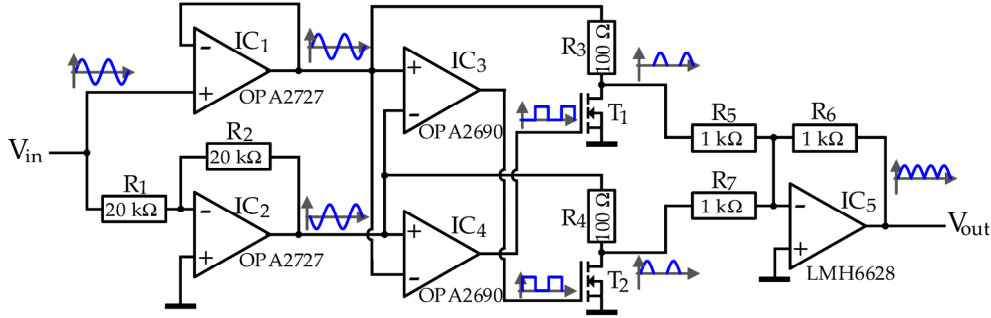

Fig. 3. Block diagram of the high-frequency full wave rectifier. At first, the input signal $V_{in}$ passes two separate paths, one to buffer ($IC_1$) the signal and the other to invert it ($IC_2$). Afterwards, the comparators $IC_3$ and $IC_4$ control the N-channel transistors $T_1$ and $T_2$ to realize two complementary half-wave rectifiers. The resulting output signals are combined to a full-wave rectified signal by the summing amplifier $IC_5$.

TABLE I
IMPLEMENTED ANALOG FILTERS FOR CHANNEL SEPARATION

| Channel | Type | Characteristic | Order | Cut-off frequencies |
|---|---|---|---|---|
| 1 | Low Pass | Butterworth | $8^{th}$ | 12 kHz |
| 2 | Band Pass | Butterworth | $6^{th}$ | 26 kHz / 34 kHz |
| 3 | Band Pass | Butterworth | $6^{th}$ | 72.5 kHz / 87.5 kHz |
| 4 | High Pass | Butterworth | $8^{th}$ | 220 kHz |

### D. Channel separation

For executing multi-channel measurements, the four channels can be separated by using different excitation current frequencies. For that purpose, every channel's internal analog oscillator, which can be connected to the current source, works with a different frequency (CH1: 10 kHz, CH2: 30 kHz, CH3: 80 kHz, CH4: 220 kHz). By using the optional analog filters of the voltage measurement inputs, the frequency components of the other channels are attenuated. The simulated filters' frequency responses are shown in Fig. 4. Additionally, the corresponding signal frequencies are plotted as dashed lines. This filter setup ensures, that the unwanted frequency components are attenuated by at least 56 dB. The channel separation may further be improved by a proper electrode setup where adjacent frequencies are allocated to maximally spaced electrodes.

The characteristics of the designed filters are shown in Table I. All filters are realized in Multiple Feedback topology using the OPA2727 (Texas Instruments, Dallas, TX, US).

### E. Software architecture and information flow

The firmware of the microcontroller is programmed in C language, using the Atmel Studio 6.0. As shown in Fig. 5 (a), its task is controlling the multiplexers and PGAs and communicating with the ADC via the SPI interface with a data rate of 5 Mbit/s. The acquired ADC data and information about the hardware status are combined into common data frames and sent via the UART interface to the USB interface chip with a baud rate of 2.5 MBaud/s. These packets are forwarded by the interface chip to the host PC, using a Full Speed USB 2.0 interface. By interpreting the USB port as a Virtual COM port, Windows programs like MATLAB or the GUI have easy access to the received data.

The GUI is programmed in C# language, using the Visual Studio 2013 (Microsoft Corporation, Redmond, WA, US). It controls the virtual COM port and can be used to configure the measurement setup. The received measurement data from all ADC channels can optionally be filtered (FIR low pass, $101^{st}$ order) and plotted in real-time. Fig. 5 shows a screenshot of the measurement GUI. The software is divided into two windows, one for real-time configuration (b) of the multiplexers and PGAs, the other for real-time data plotting (c). After a measurement procedure, the acquired raw data are stored in a comma-separated values file.

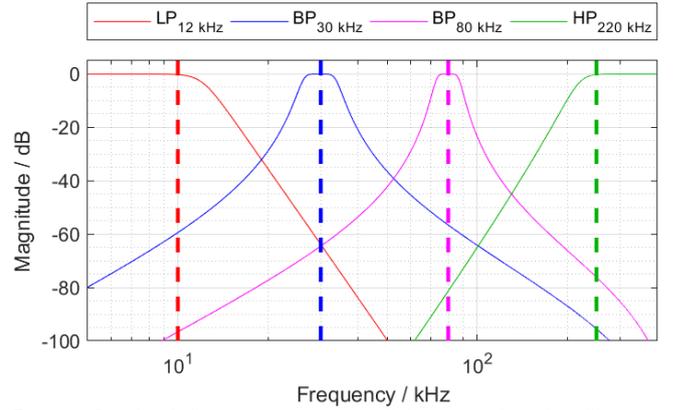

Fig. 4. Simulated frequency responses of the optional analog filters to separate the measurement channels. Additionally, the dashed lines mark the corresponding signal frequencies, which shall be decoupled from each other.

## III. SYSTEM PERFORMANCE

### A. Switched rectifier

To analyze the characteristics of the full-wave rectifier circuit, a sinusoidal test signal with a frequency of 50 kHz and an amplitude of 1.5 V is connected to a voltage input of the system. It is measured before and after rectification with a digital oscilloscope (HDO6054, Teledyne LeCroy, Chestnut Ridge, NY, US). For this measurement the oscilloscope is set to a sample rate of 250 mega samples per second (MSPS) and a resolution of 12 bits. The original sinusoid is rectified digitally and both signals are normalized in MATLAB.

In Fig. 6, both occurring signals are plotted in time (a) and frequency (b) domain. It can be seen, that there are distortions during the analog switching process at the zero crossings, which correspond to a small signal elevation in the frequency domain at the base frequency of 50 kHz. Since this frequency component's magnitude is about 60 dB lower than the DC component, it is negligible. All other differences in the frequency domain are as well negligible.

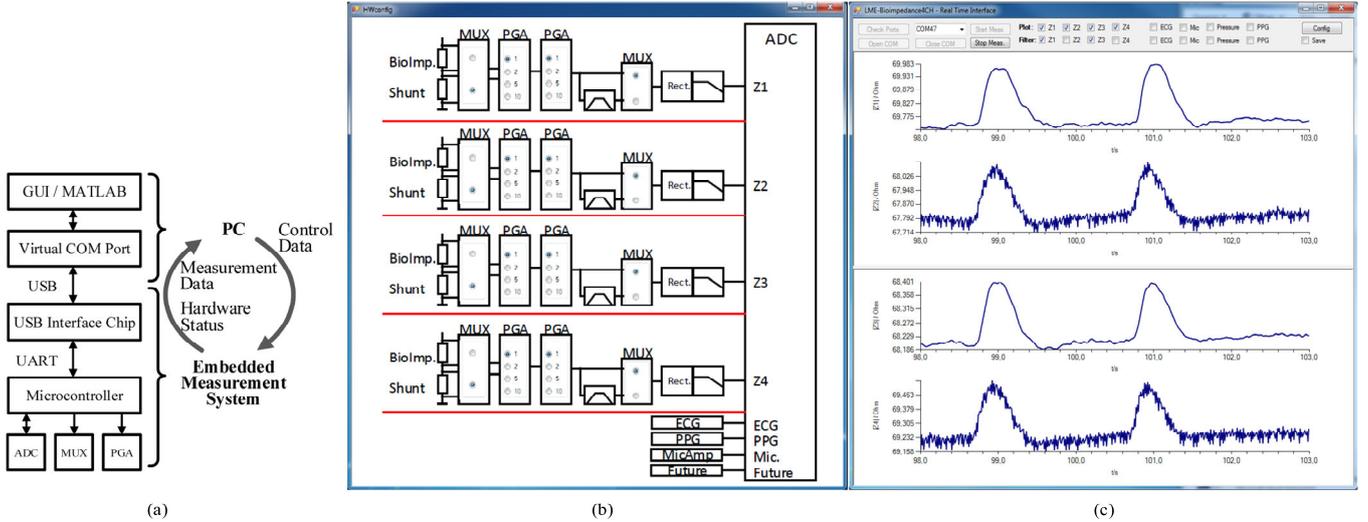

Fig. 5. Software development. (a): Software architecture and information flow between the measurement system and a host PC. (b): Screenshot of the GUI window for adjusting the system configuration. (c): Exemplary real time plots of the four impedance channels.

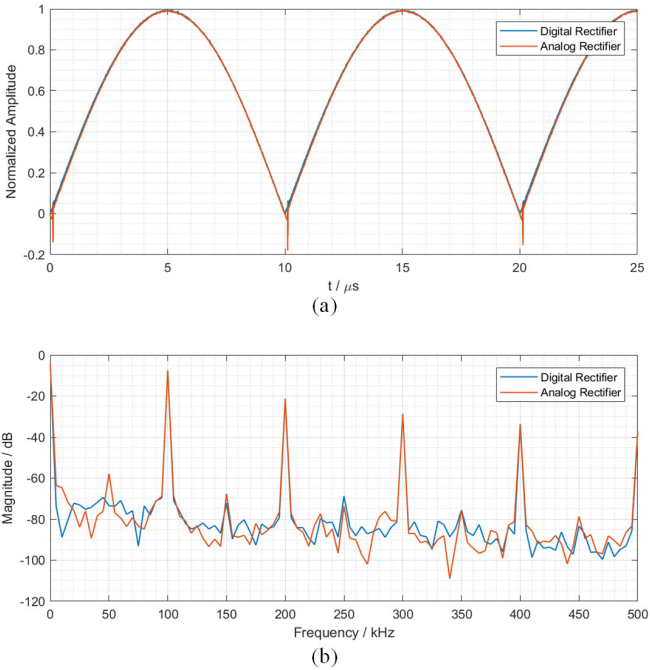

Fig. 6. Comparison between the implemented analog amplifier (red) and a digital reference rectification (blue) in time (a) and frequency (b) domain. The raw signals were acquired via an oscilloscope with a sample rate of 250 MSPS and a resolution of 12 bits for 200 µs. This leads to a frequency resolution of 5 kHz.

*B. Calibration*

In most biomedical measurement applications, just the change of the bioimpedance over time is of interest. Especially, because small changes of the electrodes' positioning lead to significant changes in the magnitude of the bioimpedance, absolute measurement values have to be interpreted carefully. Nevertheless, there are some applications like body composition measurements, in which case the absolute bioimpedance magnitude has to be measured. Therefore, a calibration is executed.

Since apart from the analog rectifier, the voltage measurement part in the block diagram in Fig. 1 consists of linear components, it is assumed, that a linear calibration is sufficient for this application. The PGAs low gain error of less than 0.04 % and gain non-linearity of less than 10 ppm leads to the acceptable simplification of calibrating the linear slope error with just one PGA configuration, but the occurring offset errors are measured once for every PGA configuration.

The usage of just one measurement circuit for both the bioimpedance and the shunt resistor allows to calibrate this single circuit by using the known shunt resistor, instead of connecting external calibration resistors.

The influence of the PGA-setup depending offset error $E_{Offset}$ (PGAconf) and the slope error $E_{slope}$, valid for all PGA setups, on the calculated impedance value $|Z_{uncalibrated}|$ is shown in the Formula below. $|V_{BI}|$ and $|V_{shunt}|$ represent the actual voltages over the bioimpedance and the shunt resistor without the influence of the measurement device.

$$|Z_{uncalibrated}| = \frac{|V_{BI}| \cdot E_{slope} + E_{Offset}(PGAconf_{BI})}{|V_{shunt}| \cdot E_{slope} + E_{Offset}(PGAconf_{shunt})} \cdot R_{shunt}$$

It can be seen, that after subtracting the offset error values from the voltage and the current measurement each, the $E_{Slope}$ values in the numerator and denominator can be cancelled out. This is only valid, if the characteristic of the device didn't change between both measurements and both slope errors can be assumed as being the same. Additionally, it has to be assumed that the excitation current is stable against impedance load changes over time.

It depends on the requirements of each measurement application, if both voltage drops have to be measured alternatingly to keep the influence of slope drifting small, or if it's for example sufficient to measure the voltage drop over the shunt resistor just once at the beginning of an impedance measurement procedure.

In this work, the four channels of the device are calibrated once. For this purpose, the offset voltage errors are determined by measuring the voltage over a short circuit for all PGA configurations. Afterwards, the slope error is determined by measuring the voltage drop over the shunt resistor, using a gain of 10, which is the highest selectable gain for the implemented shunt resistor of 69.8 Ω. For calculating both, the offset errors and the slope error, the average of 2000 measured values is taken.

TABLE II
SYSTEMATIC AND STATISTICAL ERRORS

| $R_{true}$ / Ω | 10 | 21.5 | 38.3 | 46.4 | 100 | 215 | 261 | 464 | 1000 |
|---|---|---|---|---|---|---|---|---|---|
| Gain | 100 | 50 | 25 | 20 | 10 | 5 | 4 | 2 | 1 |
| \|Max. abs. Error / mΩ\| | 0.9 | 1.7 | 2.8 | 3.2 | 7.8 | 24 | 25 | 34 | 71 |
| \|Max. rel. Error / ppm\| | 88 | 79 | 72 | 68 | 78 | 110 | 97 | 72 | 71 |
| Standard Deviation / mΩ | 0.19 | 0.42 | 0.60 | 0.97 | 1.88 | 4.81 | 6.29 | 9.48 | 14.5 |
| SNR / dB | 94.1 | 94.3 | 96.1 | 93.6 | 94.6 | 93.0 | 92.4 | 93.8 | 96.7 |

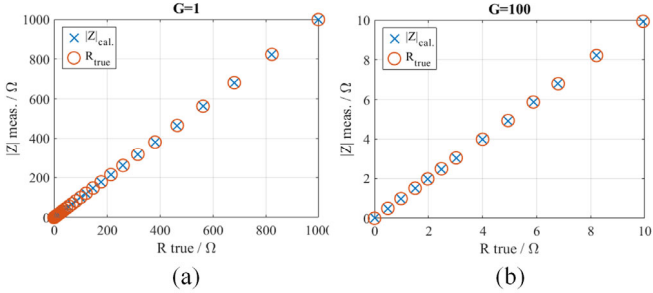

Fig. 7. Exemplary results of the measurement of known resistances for selected PGA gains of G=1 and G=100.

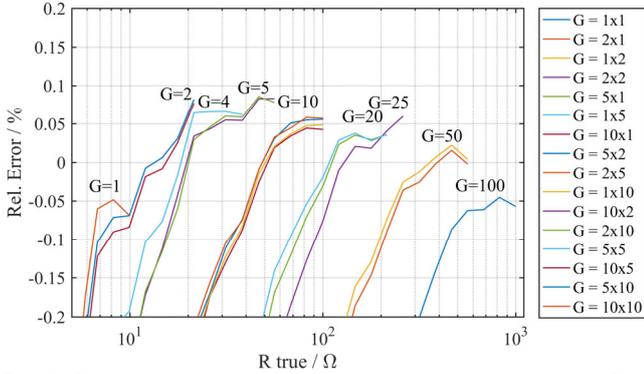

Fig. 8. Relative impedance measurement errors in a range from 6 Ω to 1000 Ω using the calibration results. Each plot represents one PGA setup. Since there are two PGAs with 4 gain stages each, all resulting 16 plots are displayed.

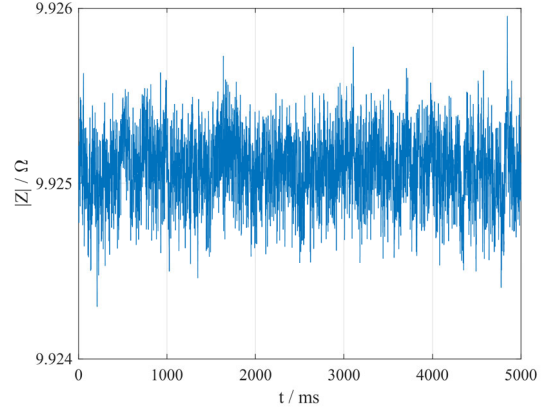

Fig. 9. Exemplary measurement result of a 9.94 Ω resistor over 5 seconds. The standard deviation in this example is 0.19 mΩ.

*C. Measurement errors*

The systematic error is determined by measuring 28 known resistors with tolerances of less than 0.1 % in the range from 6 Ω to 1000 Ω. To remove the influence of the statistical error, the resistors are measured 500 times each and the resulting values are averaged. For the measurement, coaxial cables with a length of 30 cm are used and the device has been calibrated as described before.

In Fig. 7 the impedance measurement results of two exemplary PGA setups (Gain=1, Gain=100) are illustrated. The red circles represent the true resistor values, and the blue crosses mark the measured values.

In Fig. 8 the relative measurement errors for all 16 feasible PGA combinations are shown. It can be seen that the maximum measurement error is less than 0.2 % for all measured resistors from 6 Ω to 1000 Ω, when using a suitable PGA configuration. Additionally, this plot demonstrates that the combination of PGA configurations to realize the required gain doesn't have a significant influence on the results. All four PGA combinations to set the gain to 10, for example, lead to almost the same result.

For determining the statistical errors, known resistors are measured 5000 times within 5 s each. Since the signal quality of the microcontroller's DAC is not high enough to demonstrate the precision of the remaining circuit components, an external signal generator (HP33120A, Hewlett-Packard, Palo Alto, CA, US) is used as explained before in the system architecture section. The resulting generated excitation current is 2 mA and has a frequency of 50 kHz. For this measurement, also coaxial cables with a length of 30 cm are used. In Fig. 9 an exemplary measurement of a 9.94 Ω resistor is shown. The maximum absolute statistical error in this example is 0.9 mΩ, which is about 88 ppm of the original value.

After the measurements, the standard deviations and SNR values are calculated.

In Table II the resulting measured and calculated values of the resistances are shown. The maximum absolute statistical uncertainties are in the range of milliohms and the SNR values are in the range from 92.4 dB to 96.7 dB.

For determining the long-term drift, a 100 Ω resistor is measured for a duration of 30 minutes. The mean values, averaged over 1 minute each, have a maximum drift of 5 mΩ within this time period.

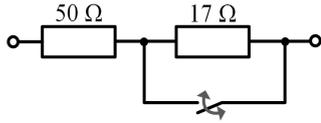

Fig. 10. Setup of two resistors for generating impedance steps from 67 Ω to 50 Ω.

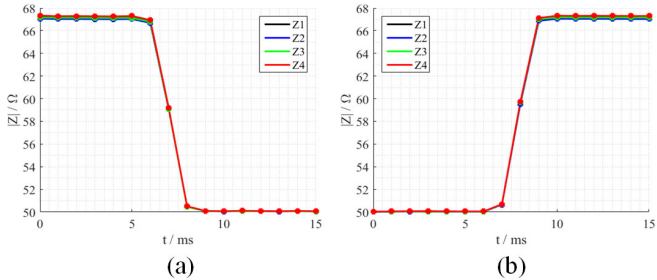

Fig. 11. Falling (a) and rising (b) step response of all four impedance channels. Just one current source was connected to the impedance in Fig. 10.

### D. Channel synchronicity

Since one of the major applications is the simultaneous measurement of bioimpedances at different positions on the subject, the synchronicity between the four channels is mandatory. To ensure meaningful results for measurements like the measurement of the pulse transient time between two channels, the time shift between the channels should be less than 1 ms. To analyze the synchronicity, a setup, as shown in Fig. 10 is used for generating immediate impedance steps.

A current of 2 mA with a frequency of 50 kHz, generated by the microcontroller's DAC, is injected into the switched impedance. All four voltage measurement channels are connected to this resistor setup, using a gain of 2. The measured falling (a) and rising (b) impedance steps of all channels are shown in Fig. 11. In both cases the rise/fall time$_{90\%}$ is about 2 ms, which is equivalent to two ADC samples. This slope is reasonable due to the step response of the analog rectifier output filter and the ADC's anti-aliasing filter.

The resolution of this measurement is limited by the sample rate of the ADC to 1 ms. The measured synchronicity is clearly below this value, which is sufficient for most applications.

### E. Frequency dependencies

Since the system is intended to detect impedance changes, the influence of the impedance's frequency on the measurement results has to be analyzed.

The flatness of the system's frequency response is especially important for bio signals, which consist of different frequency components, like the arterial pulse wave.

It's difficult to generate impedances whose magnitudes change sinusoidally over time with known frequencies to measure the frequency response of the system. Therefore, the step response, acquired as for determining the channel synchronicity before, is used. Afterwards, this step response is differentiated to get the system's impulse response which is transformed via the Fourier Transform into the frequency domain. To obtain a high frequency resolution, the step response is measured for a duration of 2 s with a sampling rate of 1000 SPS, which leads to a frequency resolution of 0.5 Hz. The resulting normalized frequency response of the system is shown in Fig. 12.

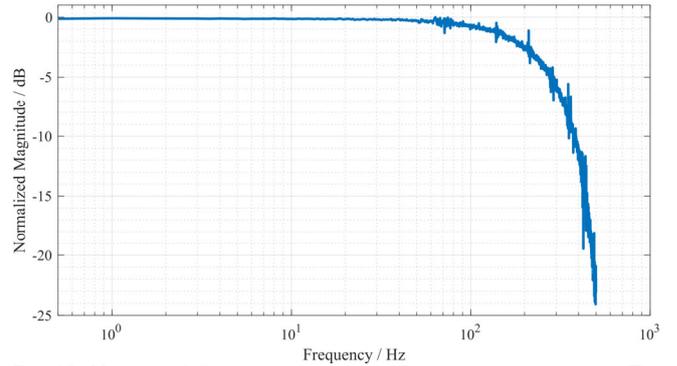

Fig. 12. Normalized frequency response of the measurement system. The transfer function is calculated via the Fourier Transform of the system's impulse response.

In a frequency range of up to 60 Hz, impedance changes are attenuated by less than 0.5 dB. This frequency range is sufficient for acquiring arterial pulse waves.

## IV. MEASUREMENT RESULTS

### A. Detection of the pulse wave at the extremities

One application of the bioimpedance measurement device is the acquisition of pulse waves at the extremities to compare their morphologies. To show the ability of measuring these minuscule impedance changes, which are caused by the blood volume changes in the arteries, a setup as shown in Fig. 13 (a) is applied to a sitting, 30 years old male subject. At the forearms, the positive electrodes are positioned at the crook of the arms and the negative electrodes are attached to the wrists, which results in a distance of 25 cm between the voltage electrodes. The electrode positions at the legs are the hollows of the knees (negative electrodes) and the ankle joints (positive electrodes). This setup leads to a distance of 35 cm between the voltage electrodes. The four bioimpedances are measured synchronously on both fore arms and both lower legs, using all four current sources and voltage measurement channels. The 2 mA excitation currents have a frequency of 50 kHz and the ADC channels digitize with a sample rate of 1 kSPS. For electrical safety purposes a medical power supply (MPU31-102, SINPRO Electronics, Pingtung City, TW) is used and the USB port is isolated (USB-GT-MED-D, Meilhaus Electronic, Alling, DE). The used standard silver chloride electrodes (Kendall H92SG, Covidien, Dublin, IE) are connected via coaxial cables to the measurement device.

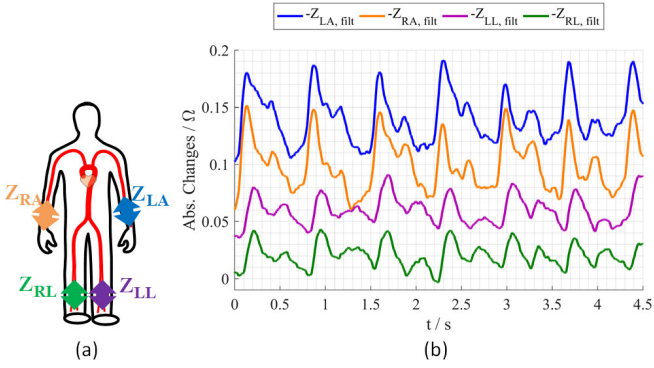

Fig. 13. Measurement setup (a) to acquire the pulse wave at all four extremities and resulting impedances over time (b). In the plot, the four impedances are filtered and for better displaying offsets are added.

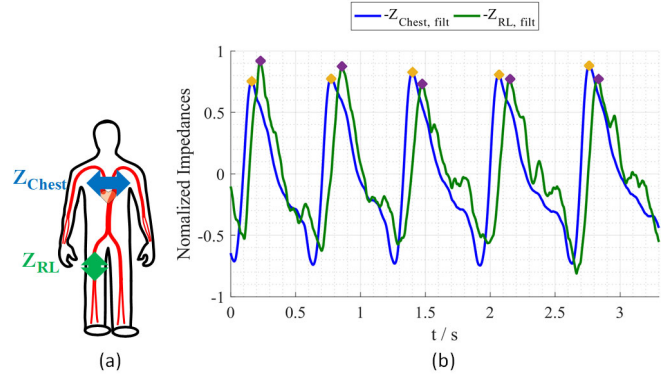

Fig. 14. Measurement setup (a) to estimate the pulse wave velocity between the aortic arch and the thigh. In the filtered and normalized measurement results (b) the transient time and change in morphology can be seen.

The measurement data are acquired for 20 seconds. Afterwards, the signals are low pass ($f_{cut-off}$ = 8 Hz) and high pass ($f_{cut-off}$ = 0.5 Hz,) filtered using 1st order zero-phase IIR filters to prevent ringing artefacts. For better visualization, offsets are added and the signals are inverted before plotting.

In the section with a duration of 4.5 s in Fig. 13 (b) it can be seen, that the impedance changes over time of all four channels have shapes like arterial pulse waves. Also the morphologies of the pulse waves from the arms look different than those from the legs. Additionally, the time delay between the pulse waves from the arms and the pulse waves from the legs are noticeable.

### B. Estimation of the pulse wave velocity in the aorta

For estimations of the arterial stiffness of the aorta, a common technique is to measure the pulse wave velocity (PWV). Since it is difficult to detect the exact starting time of the pulse wave at the heart and the arrival time of the wave at the end of the aorta by using common techniques, the pulse wave is usually acquired at other positions. These positions can be the arteria carotis (Compilor, Alam Medical, Saint Quentin Fallavier, FR), the radial artery (SphygmoCor CVMS, AtCor Medical, Itasca, IL, US) or the brachial artery (Arteriograph, TensioMed, Budapest, HU; SphygmoCor XCEL PWA, AtCor Medical, Itasca, IL, US) [28]. Instead of determining the PWV by measuring the transient time of the pulse wave between two different positions simultaneously, most of these devices execute just a single-point measurement. Afterwards, the PWV inside the aorta is estimated with the help of transfer functions. Depending on the measurement setup and the implemented algorithms, the resulting values of PWV can vary widely [28].

In Fig. 14 (a) a principle is proposed, which enables a direct determination of the PWV inside the aorta by performing two simultaneous bioimpedance measurements. The acquisition of $Z_{Chest}$ is supposed to measure the impedance changes at the position of the aortic arch for determining the starting time of the pulse wave. Therefore, the four electrodes are positioned horizontally between the 2nd and the 3rd ribs with a distance of 10 cm between the voltage electrodes. $Z_{RL}$, acquired at the upper part of the right leg (RL), contains the information about the pulse arrival time close to the end of the aorta. The positive electrodes are positioned at the popliteal fossa and the negative electrodes are positioned at the groin. No additional skin preparation has been performed. Both signals are acquired from a sitting, 30 years old male subject by using 2 mA currents with frequencies of 50 kHz for a duration of 20 s, using standard silver chloride electrodes (Kendall H92SG, Covidien, Dublin, IE) and coaxial cables. Afterwards, both signals are low pass ($f_{cut-off}$ = 6 Hz) and high pass ($f_{cut-off}$ = 1 Hz) filtered by 1st order zero-phase IIR filters to prevent ringing artefacts and normalized for better display.

In the section of the measurement in Fig. 14 (b) the time delay of the pulse wave from the right leg is depicted. Averaging of the time delays of the pulse wave peaks, as indicated with orange and purple markers, over 24 pulse waves, leads to a mean pulse transient time (PTT) of approximately 78 ms. Assuming a measured distance ($\Delta$x) of 60 cm, between the aortic arch and the measurement position at the right leg results in a mean pulse wave velocity of about 7.7 m/s, as shown in the following formula.

$$PWV_{mean} \approx \frac{\Delta x}{PTT} \approx \frac{0.6\ m}{78\ ms} \approx 7.7\ \frac{m}{s}$$

This result is surely influenced by the portion of the leg arteries and the spot of the electrodes, but it can be assumed that the determination and elimination of this influence is more reliable than the models used in conventional methods. An approach to minimize the influence of the leg arteries could be the additional determination of the PVW within the leg. Therefore, a third bioimpedance measurement could be performed at a lower leg position. The only remaining unknowns would be the exact lengths of the aorta and of the section between the aorta and the first impedance measurement at the leg. However, this unknown value influences the results of conventional methods, as well [28].

The actual advantage of this proposed measurement approach has to be analyzed and compared with the conventional methods more detailed in the future.

## V. SUMMARY AND OUTLOOK

This work describes the development of an embedded measurement system for acquiring four bioimpedances simultaneously. The focus of the device is the real-time measurement of minuscule changes in bioimpedances, caused by the pulse wave in the arteries.

The system provides four independent impedance measurement channels, consisting of voltage controlled current sources and analog voltage measurement circuits, each. Impedance measurements in the range of up to 1000 Ω can be performed by the usage of programmable gain amplifiers. The demonstrated analog rectifier circuit is able to work with signal frequencies of up to 500 kHz.

It was demonstrated, that the system has a very linear behavior and changes of impedances can be measured reliably. An SNR in a range from 92 dB to 96 dB allows to detect minuscule impedance changes in mΩ ranges.

Two exemplary measurement setups are shown to demonstrate the system's eligibility for pulse wave analyses.

In the future, the system could be improved to allow a more flexible change of the excitation current frequencies and amplitudes. Additionally, the ability of phase shift and multi-frequency measurements could be implemented, if there are interesting applications for it.

## REFERENCES

[1] A. P. Guerin et al., "Impact of Aortic Stiffness Attenuation on Survival of Patients in End-Stage Renal Failure," *Circulation*, vol. 103, no. 7, pp. 987-992, Feb. 2001.
[2] S. Laurent et al., "Expert consensus document on arterial stiffness: methodological issues and clinical applications," *Eur. Heart J.*, vol. 27, no. 21, pp. 2588-2605, Nov. 2006.
[3] S. Laurent et al., "Aortic stiffness is an independent predictor of all-cause and cardiovascular mortality in hypertensive patients," *Hypertension*, vol. 37, no. 5, pp. 1236-1241, May 2001.
[4] A. P. Avolio, M. Butlin, and A. Walsh, "Arterial blood pressure measurement and pulse wave analysis--their role in enhancing cardio-vascular assessment," *Physiol. Meas.*, vol. 31, no. 1, pp. R1-R47, Jan. 2010.
[5] S. Wassertheurer et al., "A new oscillometric method for pulse wave analysis: comparison with a common tonometric method," *J Hum Hypertens.*, vol. 24, no. 8, pp. 498-504, Aug. 2010.
[6] L. Luzardo et al., "24-h ambulatory recording of aortic pulse wave velocity and central systolic augmentation: a feasibility study," *Hypertens Res.*, vol. 35, no. 10, pp. 980-987, Oct. 2012.
[7] M. W. Rajzer et al., "Comparison of aortic pulse wave velocity measured by three techniques: Complior, SphygmoCor and Arteriograph," *J Hypertens.*, vol. 26, no. 10, pp. 2001-2007, Oct. 2008.
[8] C. Vlachopoulos, K. Aznaouridis, and C. Stefanadis, "Prediction of cardiovascular events and all-cause mortality with arterial stiffness: a systematic review and meta-analysis," *J. Am. Coll. Cardiol.*, vol. 55, no. 13, pp. 1318-1327, Mar. 2010.
[9] W. W. Nichols, "Effects of arterial stiffness, pulse wave velocity, and wave reflections on the central aortic pressure waveform," *J Clin Hypertens.*, vol. 10, no. 4, pp. 295-303, Apr. 2008.
[10] M. Cecelja, and P. Chowienczyk, "Role of arterial stiffness in cardiovascular disease," *JRSM Cardiovasc Dis.*, vol. 1, no. 4, pp. 1-10, Jul. 2012.
[11] J. L. Cavalcante, J. A. Lima, A. Redheuil, and M. H. Al-Mallah, "Aortic stiffness: current understanding and future directions," *J Am Coll Cardiol.*, vol. 47, no. 14, pp. 1511-1522, Apr. 2011.
[12] G. F. Mitchell, "Arterial Stiffness and Wave Reflection: Biomarkers of Cardiovascular Risk," *Artery Res.*, vol. 3, no. 2, pp. 56-64, Jun. 2009.
[13] S. Munir et al., "Peripheral augmentation index defines the relationship between central and peripheral pulse pressure," *Hypertension*, vol. 51, no. 1, pp. 112-118, Jan. 2008.
[14] S. Kaufmann, A. Malhotra, G. Ardelt, and M. Ryschka, "A high accuracy broadband measurement system for time resolved complex bioimpedance measurements," *Physiol Meas.*, vol. 35, no. 6, pp. 1163-1180, Jun. 2014.
[15] P. S. Luna-Lozano, and R. Pallàs-Areny, "Heart rate detection from impedance plethysmography based on concealed capacitive electrodes," in *Proc. XIX IMEKO World Congress*, Sep. 2009, pp. 1701-1706.
[16] R. Kusche, T. D. Adornetto, P. Klimach, and M. Ryschka, "A Bioimpedance Measurement System for Pulse Wave Analysis," in *Proc. 8th Int. Workshop on Impedance Spectroscopy*, 2015.
[17] S. Grimnes, and O. G. Martinsen, *Bioelectricity and Bioimpedance Basics* (2nd ed.), New York, Academic Press, 2008.
[18] C. Gabriel, A. Peyman, and E. H. Grant, "Electrical conductivity of tissue at frequencies below 1 MHz," Phys. Med. Biol., vol. 54, pp. 4863-4878, Jul. 2009.
[19] Y. Yang et al., "Design and preliminary evaluation of a portable device for the measurement of bioimpedance spectroscopy," *Physiol Meas.*, vol. 27, no. 12, pp. 1293-1310, Dec. 2006.
[20] T. I. Oh et al., "A fully parallel multi-frequency EIT system with flexible electrode configuration: KHU Mark2," *Physiol Meas.*, vol. 32, no. 7, pp. 835-849, Jul. 2011.
[21] G. J. Saulnier et al., "An electrical impedance spectroscopy system for breast cancer detection," in *Conf Proc IEEE Eng Med Biol Soc.*, 2007.
[22] R. Kusche et al., "A FPGA-Based Broadband EIT System for Complex Bioimpedance Measurements - Design and Performance Estimation," *Electronics*, vol. 4, no. 3, pp. 507-525, Jul. 2015.
[23] S. Khan, P. Manwaring, A. Borsic, and R. Halter, "FPGA-based voltage and current dual drive system for high frame rate electrical impedance tomography," *IEEE Trans Med Imaging.*, vol. 34, no. 4, pp. 888-901, Apr. 2015.
[24] S. Kaufmann et al., "Multi-frequency Electrical Impedance Tomography for Intracranial Applications," in *Proc. World Congress on Medical Physics and Biomedical Engineering*, 2012.
[25] X. Zhao, S. Kaufmann, and M. Ryschka, "A comparison of different multi-frequency Current Sources for Impedance Spectroscopy," in *Proc. 5th Int. Workshop on Impedance Spectroscopy*, 2012.
[26] Texas Instruments, "ADS131E0x 4-, 6-, and 8-Channel, 24-Bit, Simultaneously-Sampling, Delta-Sigma ADC," ADS131E06 datasheet, Jun. 2012 [Revised Jan. 2017]
[27] B. Sanchez et al., "A new measuring and identification approach for time-varying bioimpedance using multisine electrical impedance spectroscopy," *Physiol Meas.*, vol. 34, no. 3, pp. 339-357, Mar. 2013.
[28] M. W. Rajzer et al., "Comparison of aortic pulse wave velocity measured by three techniques: Complior, SphygmoCor and Arteriograph," *J. Hypertens.*, vol. 26, no. 10, pp. 2001-2007, Oct. 2008.

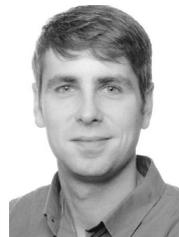
**Roman Kusche** received the B.Sc. in electrical engineering from Lübeck University of Applied Sciences, Lübeck, Germany, in 2013, and the M.Sc. from the Hamburg University of Applied Sciences, Hamburg, Germany, in 2014. He is currently a research associate with the Laboratory of Medical Electronics of the Lübeck University of Applied Sciences. His research focuses on the development of novel biomedical measurement methods and the related medical electronic devices.

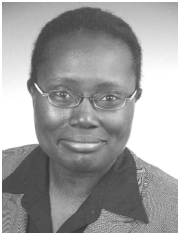
**Paula Klimach** earned a Master of Science in biomedical engineering from Lübeck University of Applied Sciences. She has worked in industry, from 2000 to 2002 as a software engineer and later as a patent examiner, from 2002 to 2006, for the United States Patent Office. Currently she is working as a research assistant at the Lübeck University of Applied Sciences.

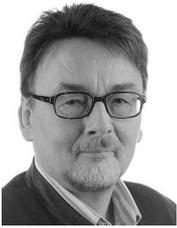
**Martin Ryschka** received a Ph.D. in physics in 1982 from Christian-Albrechts-Universität Kiel, Germany. He worked, in industry, as an R&D scientist in the medical electronic devices development field. From 1994 until 2011, he worked as the managing director of CogniMed GmbH, Germany. He has been in the position of professor for electrical engineering, since 1995, at Lübeck University of Applied Sciences.